\documentclass[runningheads]{llncs}
\usepackage[T1]{fontenc}
\usepackage{graphicx}
\usepackage{amsfonts}
\usepackage{amsmath}
\usepackage{algorithm}
\usepackage{algpseudocode}
\usepackage{censor,caption}

\begin{document}
\title{On Instabilities of Unsupervised Denoising Diffusion Models in Magnetic Resonance Imaging Reconstruction}
%
%
\author{
Tianyu Han\inst{1} \and 
Sven Nebelung\inst{1} \and 
Firas Khader\inst{1} \and 
Jakob Nikolas Kather\inst{2,3,4} \and 
Daniel Truhn\inst{1} 
}
\authorrunning{T. Han et al.}
%
\institute{
Department of Diagnostic and Interventional Radiology, University Hospital
Aachen, Aachen, Germany \and 
Else Kroener Fresenius Center for Digital Health, Medical Faculty Carl
Gustav Carus, Technische Universität Dresden, Dresden, Germany \and
Department of Medicine I, University Hospital Dresden, Dresden, Germany \and
Medical Oncology, National Center for Tumor Diseases (NCT), University Hospital Heidelberg, Heidelberg, Germany
}
\maketitle              
\begin{abstract}
Denoising diffusion models offer a promising approach to accelerating magnetic resonance imaging (MRI) and producing diagnostic-level images in an unsupervised manner. 
However, our study demonstrates that even tiny worst-case potential perturbations transferred from a surrogate model can cause these models to generate fake tissue structures that may mislead clinicians. 
The transferability of such worst-case perturbations indicates that the robustness of image reconstruction may be compromised due to MR system imperfections or other sources of noise.
Moreover, at larger perturbation strengths, diffusion models exhibit Gaussian noise-like artifacts that are distinct from those observed in supervised models and are more challenging to detect. 
Our results highlight the vulnerability of current state-of-the-art diffusion-based reconstruction models to possible worst-case perturbations and underscore the need for further research to improve their robustness and reliability in clinical settings. 

\end{abstract}
\section{Introduction}

Magnetic Resonance Imaging (MRI) is essential for medical diagnostics, especially for brain diseases, due to its detailed, non-invasive imaging capabilities. 
However, MRI faces challenges like long acquisition times and high sensitivity to motion. 
Recent advancements, particularly denoising diffusion models, promise to accelerate MRI by reconstructing high-quality images from undersampled data. 
Unlike traditional methods, these models can operate without paired training data. 
However, our study reveals a critical vulnerability: susceptibility to minimal worst-case perturbations, leading to significant inaccuracies in reconstructed images. 
Our research explores the robustness of diffusion models in MRI reconstruction, investigating adversarial perturbations and proposing strategies to enhance resilience. 
We aim to advance reliable diffusion models in clinical settings.


\section{Related Works}
\subsection{DL-based end-to-end solution}
Model-based image reconstruction methods consider imaging systems as a linear operator $\textbf{A}$ that maps anatomical ground truth to the signal domain. 
Specifically, a noisy observation $\textbf{y}$ given by $m$ sparse measurements can be defined as 
\begin{equation}
    \textbf{y} \in \mathbb{R}^m = \textbf{Ax} + \epsilon,
    \label{equ:forward}
\end{equation}
where $\textbf{x}\in \mathbb{R}^n$ is the unknown, $\textbf{A} \in \mathbb{R}^{m\times n}$ ($n > m$) denotes an operator randomly samples k-space data, and $\epsilon \in \mathbb{R}^m$ is the measurement noise. 
When using standard Cartesian acquisition, we can factorize operator $\textbf{A}$ into an operator $\mathcal{P}(\Lambda)$ and an invertible matrix $T \in \mathbb{R}^{n\times n}$ that corresponds to Fourier transform: $\textbf{A} = \mathcal{P}(\Lambda) T$.
K-space lines are selected during acquisition if $\Lambda_{ii} = 1$ in $\mathcal{P}(\Lambda)$, where $\Lambda \in \{ 0, 1 \}^{n\times n}$ is a diagonal matrix with $tr(\Lambda) = m$.

Given the rise of DL, it's a natural flow to approximate the inverse model using convolutional neural networks (CNNs).
In the CNN formulation, one force $\textbf{x}$ to be well-approximated by the CNN reconstruction by using the following objective:
\begin{equation}
    \underset{\theta}{\min.} \; \lambda \left \| \textbf{x} - f_\text{cnn}(\textbf{A}^H \textbf{y} \, | \, \theta) \right \|_2^2 + \left \| \textbf{Ax} - \textbf{y} \, \vphantom{f_\text{cnn}(\textbf{A}^H \textbf{y} \, | \, \theta)}\right \|_2^2.
    \label{equ:cnn_cs}
\end{equation}
Here, we denote $f_\text{cnn}(\cdot \, | \, \theta)$ as a CNN parameterized by $\theta$.
The CNN reconstruction can be considered as resolving a de-aliasing issue in the spatial domain because $\textbf{A}^H \textbf{y}$ is severely affected by aliasing from sub-Nyquist sampling. 
However, the performance of directly optimizing Equation \ref{equ:cnn_cs} is inadequate since the CNN reconstruction and the data fidelity are optimized independently. 
The CNN is specifically trained to reconstruct the sequence without knowing the prior details of the obtained data in k-space because it works purely in the image domain.

\subsection{Bayesian image reconstruction}
In the Bayesian picture, the MRI measurement $\textbf{y}$ and tissue signal $\textbf{x}$ are coupled by a measurement distribution in this probabilistic formulation: $p(\textbf{y} \, | \, \textbf{x}) = q_\epsilon(\textbf{y} - \textbf{Ax})$,
where $q_\epsilon(\cdot)$ stands for the noise distribution.
The conditional distribution $p(\textbf{y} \, | \, \textbf{x})$ represents a forward process of measuring $\textbf{y}$ from $\textbf{x}$, which is also described by the linear forward model (Equation \ref{equ:forward}).  
The reconstruction problem is then viewed as drawing samples from the posterior distribution $p(\textbf{x} \, | \, \textbf{y})$. 
In general, we can obtain such a posterior using Bayes' theorem: $p(\textbf{x} \, | \, \textbf{y}) = p(\textbf{y} \, | \, \textbf{x}) p(\textbf{x}) / p(\textbf{y})$.
The following Bayes' rule for score functions results from taking gradients concerning $x$ on both sides of this expression: $\nabla_\textbf{x} \log p(\textbf{x} \, | \, \textbf{y}) = \nabla_\textbf{x} \log p(\textbf{y} \, | \, \textbf{x}) + \nabla_\textbf{x} \log p(\textbf{x})$. Note, the data prior term $\nabla_\textbf{x} \log p(\textbf{x})$ can be efficiently estimated by a denoising diffusion model \cite{ho2020denoising,song2020score}. 
Incorporating measured observation $\nabla_\textbf{x} \log p(\textbf{y} \, | \, \textbf{x})$ into our system is an essential step that transforms an unguided diffusion into a conditional one. 
In general, one can add such a correction term to unconditional diffusion steps either via approximating the likelihood gradient or directly performing closed-form data consistency.

\subsection{Reconstructing MRI using diffusion prior and posterior sampling}
Given a time-dependent diffusion-based model $s_{\theta^\ast}(\textbf{x}_t, t)$ that has been trained to approximate the data score function via a diffusion process $\{\textbf{x}_t\}_{t=0}^T$ that was produced by perturbing $\textbf{x}$ with an SDE.
The procedure of unconditional sampling chooses a series of time steps and iterates by
$\hat{x}_{t_{i-1}} = h(\hat{x}_{t_i}, \, z_i, \, s_{\theta^\ast}(\hat{\textbf{x}}_t, t_i))$,
where function $h$ represents an SDE solver and $z_i \sim \mathcal{N}(0, \textit{I})$.   
In MRI reconstruction, function $h$ needs an additional step $k$ prepended to itself to impose the constraint implied by measurements, resulting in
\begin{equation}
    \begin{aligned}
        \hat{x}_{t_{i}}' &= k(\hat{x}_{t_i}, \, \hat{y}_{t_i}, \, \lambda) \\
        \hat{x}_{t_{i-1}} &= h(\hat{x}_{t_i}', \, z_i, \, s_{\theta^\ast}(\hat{\textbf{x}}_t, t_i)).
    \end{aligned}
    \label{equ:sde_observe}
\end{equation}
The above $k$ function interacts with measured k-space entries and thus should follow an image fidelity objective and a k-space consistency objective:  
\begin{equation}
    \hat{x}_t' = T^{-1}\left[ \lambda \Lambda P^{-1}(\Lambda)\hat{y}_t + (1-\lambda)\Lambda T \hat{x}_t + (\textit{I}-\Lambda)T \hat{x}_t \right].
    \label{equ:song_dc}
\end{equation}

\section{Worst-Case Instabilities in MRI reconstruction}

\begin{figure}
    \centering
    \includegraphics[trim=0 0 160 0, clip, width=\textwidth]{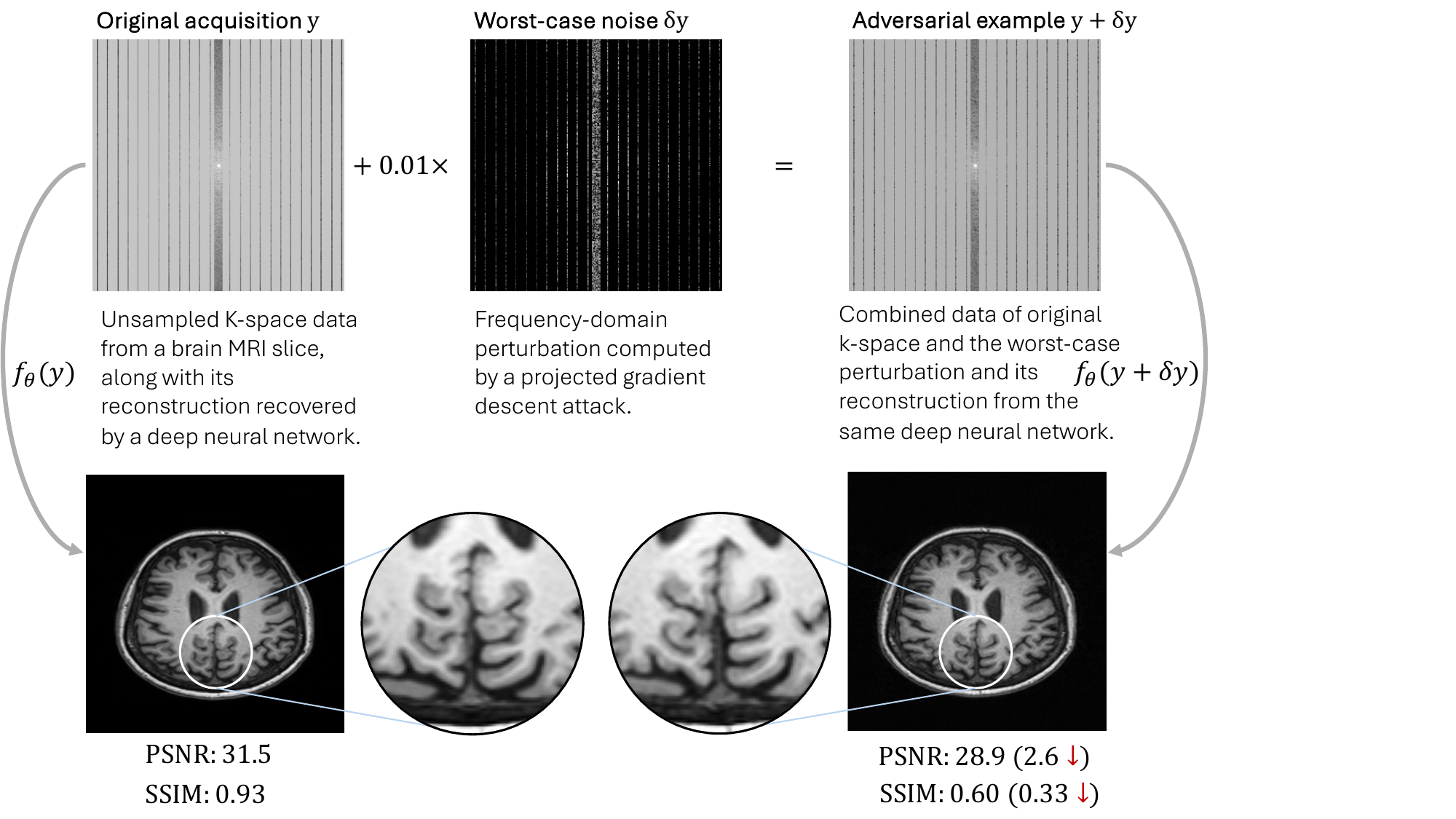}
    \caption{
      MRI reconstruction can be vulnerable to worst-case perturbations, which add noise to the original k-space signal and manipulate the reconstruction process of undersampled data. 
      The resulting reconstructions can show false gray matter structures that are difficult for humans to detect (see zoomed-in plots).
      }
    \label{fig:fig1}
  \end{figure}

As demonstrated in Fig. \ref{fig:fig1}, our worst-case noise is designed to be small in k-space but can induce a significant mismatch in the slice recovered by the undersampled version of the perturbed ground truth. 
Let $f$: $\mathbb{R}^m \rightarrow \mathbb{R}^n$ be a trained neural network $f$ mapping an undersampled measurement to an image. 
Finding an adversarial direction $\delta \in \mathbb{R}^m$ in the measurement domain can be viewed as solving an optimization with the following objective: 
\begin{equation}
  \begin{aligned}
    \underset{\delta: \left\| \delta \right\| \leq \epsilon}{\max} \mathbb{E}_\mathbf{y} \left[ \left\| f(\mathbf{y}; \theta) - f(\mathbf{y} + \delta; \theta) \right\|_2^2 \right].
  \end{aligned}
  \label{equ:adv}
\end{equation}
Here, we confine allowed perturbation sets to be a hypersphere $l_2$ ball around any $\mathbf{y}$ with a norm $\epsilon$, i.e., $\left\| \delta \right\|_2 \leq \epsilon \left\| \mathbf{y} \right\|_2$.
Following \cite{darestani2021measuring}, a projected gradient descent (PGD) method \cite{madry2017towards} was used to maximize the objective in Equation \ref{equ:adv}. 

Both white- and black-box perturbations were evaluated in this study.
In a white-box scenario, we have full access to the target reconstruction model $\theta$ and can perform gradient-based PGD attacks on its reconstruction. 
Second, in the black-box setting, we tested the success rate of transferring the above perturbations to the remaining models, e.g., diffusion-based reconstructions. 

\begin{algorithm}
    \caption{Worst-case perturbation in k-space}\label{alg:algo}
    \begin{algorithmic}
    \Require A model with its current parameter $\theta$: $f_\theta$; Loss function: $L = -L_{2}$; Partially measured k-space: $ksp$; Acquisition mask: $M$
    \Require Adam Optimizer $Opt$; Perturbation strength: $\epsilon$; Number of iterations: $T$; Learning rate: $\alpha$; Constant $c = 1e4$
    \State Initialize perturbation: $\delta \gets \delta_{r} + i \cdot \delta_{i}$, where $\delta_{r}, \delta_{i} \sim \mathcal{N}(0, \textit{I})$
    \State $\delta  \gets \delta \times \frac{\| ksp \|}{\| \delta \| \times c} $
    \State Initialize Optimizer: $Opt \gets \texttt{Adam}(\texttt{param}=[\delta], \texttt{lr}=\alpha)$
    \State Get standard reconstruction: $x \gets f_\theta(ksp \odot M, M)$
    \For{\texttt{i in range(T)}}
    \State $\hat{x} \gets f_\theta((ksp + \delta) \odot M, M)$
    \State $Opt.\texttt{zero\_grad()}$
    \State $loss \gets L(x, \hat{x})$
    \State $loss.\texttt{backward()}$

    \If{$\| \delta \| > \epsilon$}
        \State $\delta \gets \delta \times \frac{\| ksp|\|}{\| \delta \|} \cdot \epsilon$
    \EndIf
    \EndFor
    \State \Return $\delta$
    \end{algorithmic}
\end{algorithm}

\section{Experiments and Results}

\subsection{Data}
All neuroimaging data were employed from the Alzheimer's Disease Neuroimaging Initiative (ADNI), a multicenter, longitudinal study of 2,463 participants between the ages of 55 and 90 who had or were at high risk of developing dementia and Alzheimer's disease.
Each participant has gone through $T_1$ weighted structure MRI measured by Magnetization Prepared RApid Gradient Echo (MP-RAGE) \cite{mugler1990three}.
For our study, we utilized 80\% of the dataset, corresponding to 1,970 participants and 13,651 scans, for training, and randomly selected 108 scans from the remaining 20\% of ADNI participants (493 individuals) for testing. 
Using 80\% of the dataset for training and validation follows a standard 80-20 train-validation split, ensuring robust model evaluation.

\subsection{Supervised baselines}
CNNs have established a new state-of-the-art MRI reconstruction, vastly beyond the traditional baselines. 
One typical approach utilizes auto-encoder architectures, such as U-Net \cite{ronneberger2015u,zbontar2018fastmri}, which solves the medical inverse problem in an end-to-end fashion. 
We selected a Unet-based baseline (ResUnet++) as it is the most widely used CNN backbone in MRI image reconstruction.
In experiments, we trained a ResUnet++ \cite{jha2019resunet++} model with a batch size of 16 and a learning rate of 0.001, using 50 epochs on ADNI training set.
Another branch of models such as ADMM \cite{sun2016deep} and i-RIM \cite{putzky2019rim,putzky2019invert} generalizes the idea of iterative compressed sensing reconstruction that unrolls the data-flow graph via a cascade of neural networks. 
We selected i-RIM model due to its superior performance in vairous MRI reconstruction challenges, especially, the FastMRI challenge \cite{putzky2019rim,knoll2020advancing}.
In our experiments, the i-RIM model was trained using a batch size of 4 and a learning rate of 0.001 with 50 epochs. 
We trained both supervised models with an acceleration factor of 8. 

\subsection{Denoising diffusion reconstruction}
Our training method is similar to Song et al \cite{song2020score}. 
Since the Predictor-Corrector (PC) sampler has generally higher performance for VE-SDEs, we employ it here in place of the numerical SDE solver to generate samples. 
The corrector in this PC sampler follows Langevin dynamics that solely rely on the scores, whereas the predictor refers to a numerical solver for the reverse-time SDE.
We perform an additional data consistency \cite{qin2018convolutional,song2020score} step to adapt the PC sampler for solving inverse problems. 
1,000 noise scales and 1 step of the Langevin correction for each noise scale were selected, resulting in a total of 2,000 steps of score model evaluation in the PC sampler.
Besides, the signal-to-noise ratio (SNR) $\eta$ controls the step size $\epsilon$ in Langevin dynamics.
In our setting, $\eta$ was set to 0.517 and the $\lambda$ in the data consistency operation was set to 1.0.

\begin{figure}
  \centering
  \includegraphics[trim=150 1350 100 0, clip, width=\textwidth]{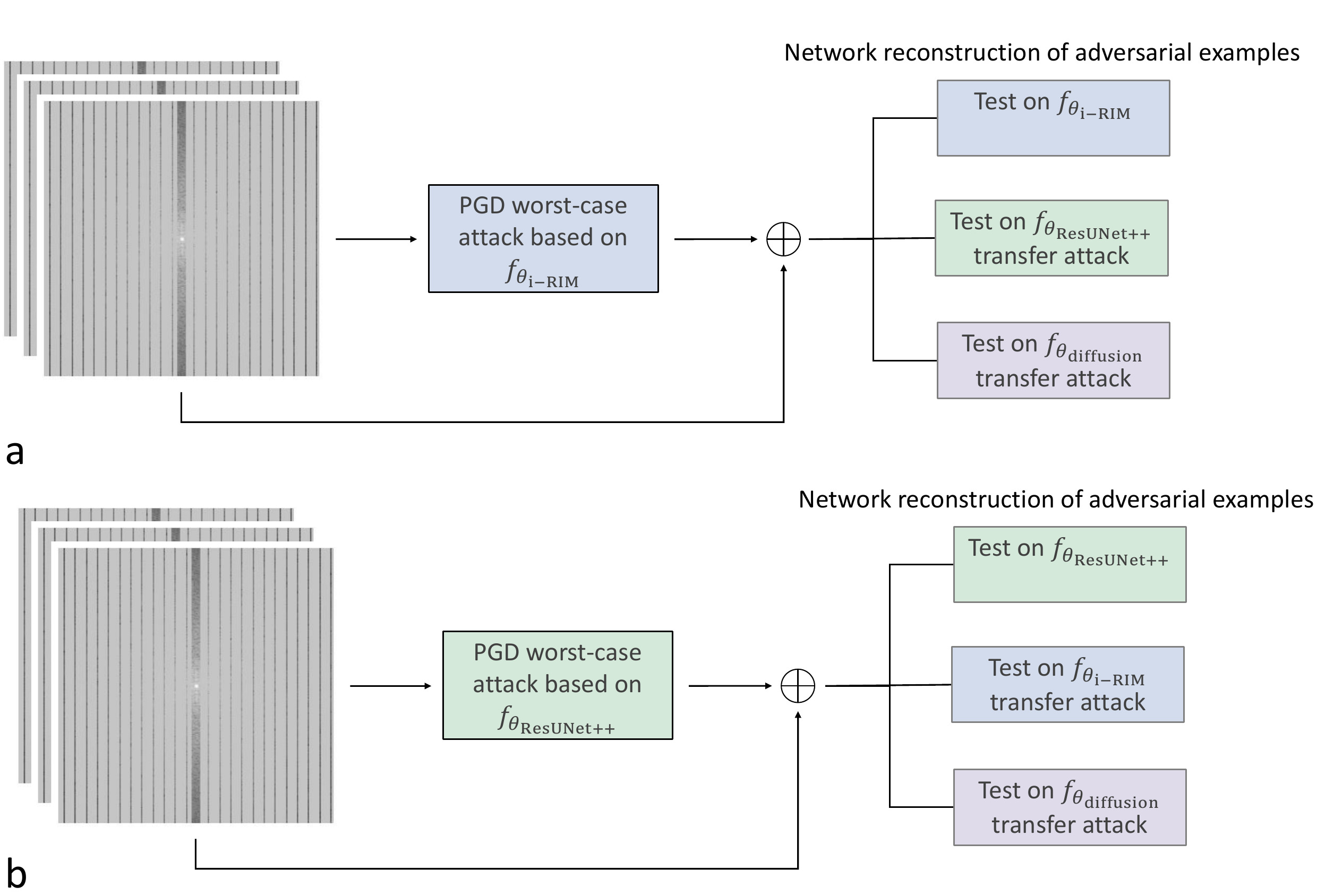}
  \caption{
    We designed experiments to evaluate the susceptibility of trained i-RIM and ResUnet++ models to white- and black-box attacks (\textbf{a} and \textbf{b}). 
  }
  \label{fig:fig2}
\end{figure}

\subsection{Experimental design}
First, we trained supervised and unsupervised models on reconstruction tasks in the training cohort and evaluated the performance in the test cohort with three acceleration schemes.
We demonstrated diffusion models deliver comparable performance to state-of-the-art supervised models while showing significantly better generalization to unknown acquisition processes. 
Then, as demonstrated in Fig. \ref{fig:fig2}, we assessed the robustness of the trained models against white- and black-box adversarial perturbations.  
Both scenarios are important as a white-box attack corresponds to malfunctions of an internal MRI system, while black-box perturbations reveal vulnerabilities of current deep-learning reconstruction towards possible external adversarial interferences. 

\subsection{Quantitative and visual evaluation}

\begin{figure}
  \centering
  \includegraphics[trim=0 140 120 0, clip, width=\textwidth]{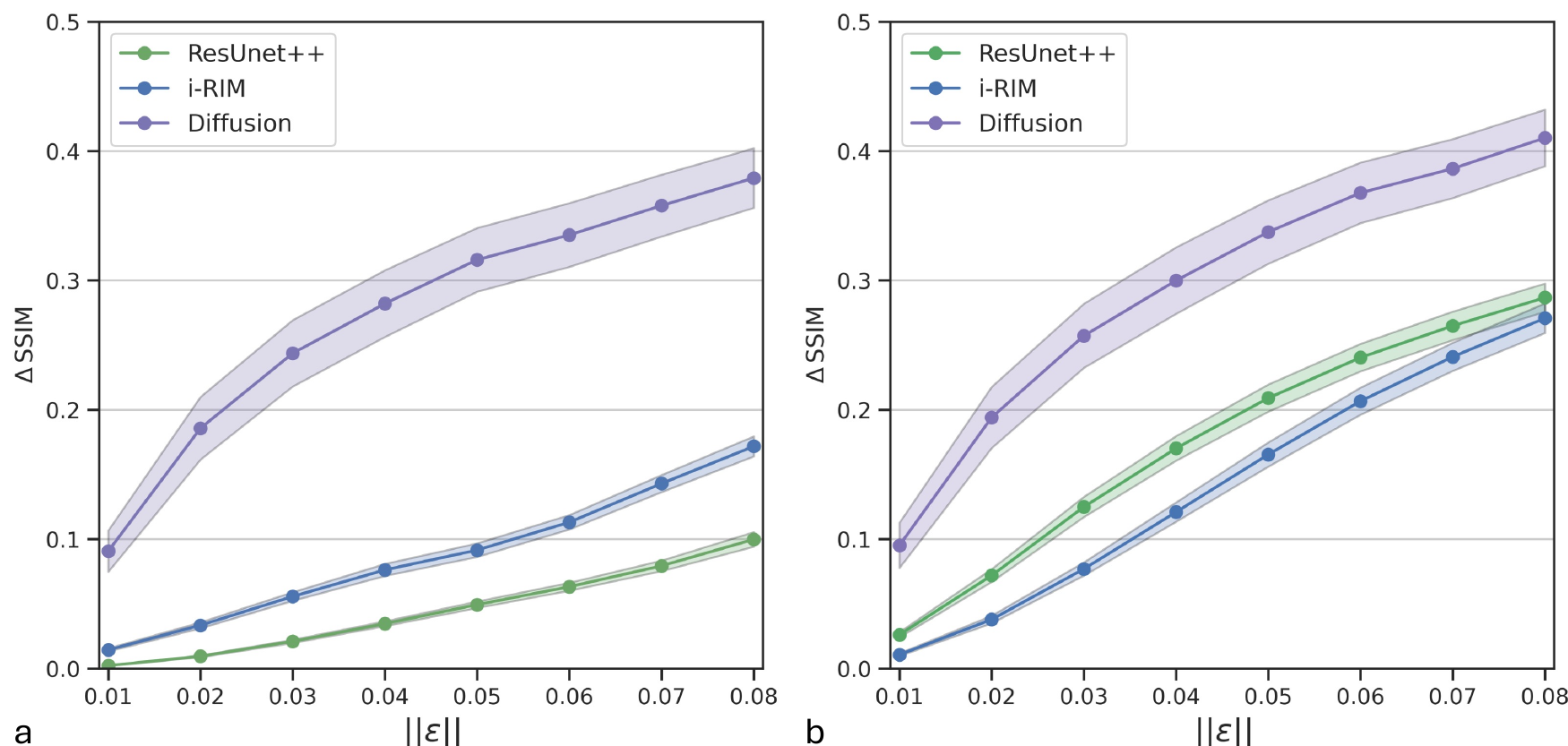}
  \caption{
    We visualized the impact of perturbation amplitude on model performance, measured by the $\Delta$SSIM metric. 
    Subplot (a) shows that all models experienced a drastic drop in SSIM as the perturbation amplitude increased using worst-case perturbations generated by i-RIM. 
    Similar findings were observed with adversarial perturbations via the ResUnet model, in (b).
  }
  \label{fig:fig3}
\end{figure}

\subsubsection{Worst-case instabilities of supervised models.}

The results of our experiments are presented in Fig. \ref{fig:fig3}, which shows the SSIM loss as a function of perturbation strength. 
Subplots c and d of Fig. \ref{fig:fig3} show the white-box perturbations \cite{han2021advancing} obtained by attacking i-RIM and ResUnet++ models, respectively. 
We found that both i-RIM (the blue line in subplot (a)) and ResUnet++ (the green line in subplot (b)) are unstable and can be easily biased by tiny adversarial perturbations.

\subsubsection{Worst-case transferability to diffusion models.}

\begin{figure}
  \centering
  \includegraphics[trim=150 1270 540 560, clip, width=\textwidth]{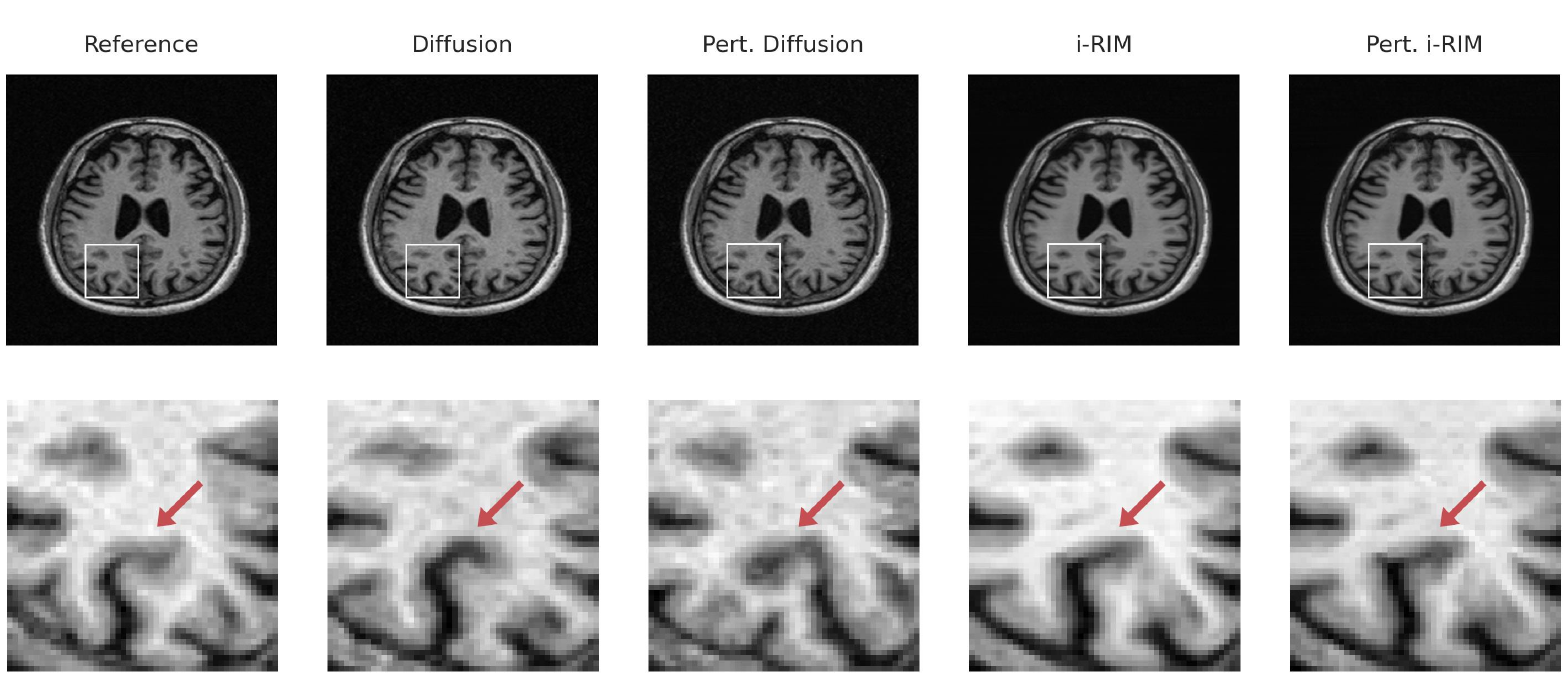}
  \caption{
    To demonstrate, we crafted worst-case black-box perturbations \cite{ghaffari2022adversarial} using an independent ResUnet++ model and applied them to unsupervised diffusion reconstruction and supervised i-RIM.
    The application of worst-case inference to unsupervised reconstruction can create misleading artifacts in brain tissue, which can be seen as red arrows in the subplot below. 
  }
  \label{fig:fig4}
\end{figure}

Empirical evidence for the transferability of adversarial examples has been investigated in classification applications, but rarely demonstrated in regression tasks such as image reconstruction. 
In Fig. \ref{fig:fig2}a, we explore a black-box scenario where adversarial perturbations are generated against a surrogate model, such as an i-RIM, rather than the actual models used for reconstruction, such as diffusion. 
Similar to supervised models, as shown in Figures \ref{fig:fig3}, the unsupervised diffusion model is also susceptible to worst-case distribution shifts in the form of adversarial perturbations. 
Even with a minimal adversarial perturbation of $\epsilon = 0.01$ crafted from ResUnet++ parameters, the diffusion model distorts the gray matter structure in its reconstruction (Fig. \ref{fig:fig4}). 
This result is significant because diffusion models are trained similarly to denoisers, as evidenced by the appearance of a Gaussian-noise-like artifact when we adversarially perturb them (Fig. \ref{fig:fig4}).

\section{Conclusion}

In summary, our study highlights the vulnerability of both state-of-the-art supervised models and unsupervised diffusion models to adversarial perturbations from the MRI signal domain. 
We found that worst-case perturbations can effectively transfer between independently trained regression models, similar to the transferability observed in classification tasks. 
While diffusion models are generally robust against anatomical and test-time distribution shifts, our findings indicate that even tiny adversarial perturbations can cause these models to generate fake tissue structures that may mislead clinicians. 
Furthermore, at larger perturbation amplitudes, diffusion models exhibit noise-like artifacts that are distinct from those observed in supervised models and may be more difficult for clinicians to detect. 

We hypothesize that the main reason for this vulnerability is due to the perturbed K-space misleads the reverse iterative diffusion process, creating nonphysical artifacts. 
Classical regularization techniques like total variance regularization might offer better robustness in such scenarios.

\begin{credits}
\subsubsection{\discintname}
The authors have no competing interests to declare that are
relevant to the content of this article.
\end{credits}

%
%
%
%
%
%
\clearpage
\bibliographystyle{splncs04}
\bibliography{Paper-1977}
\clearpage
\section*{Supplementary Information}
\setcounter{figure}{0}
\renewcommand{\thefigure}{S\arabic{figure}}

\begin{figure}[h!]
  \centering
  \includegraphics[trim=0 160 160 0, clip, width=\textwidth]{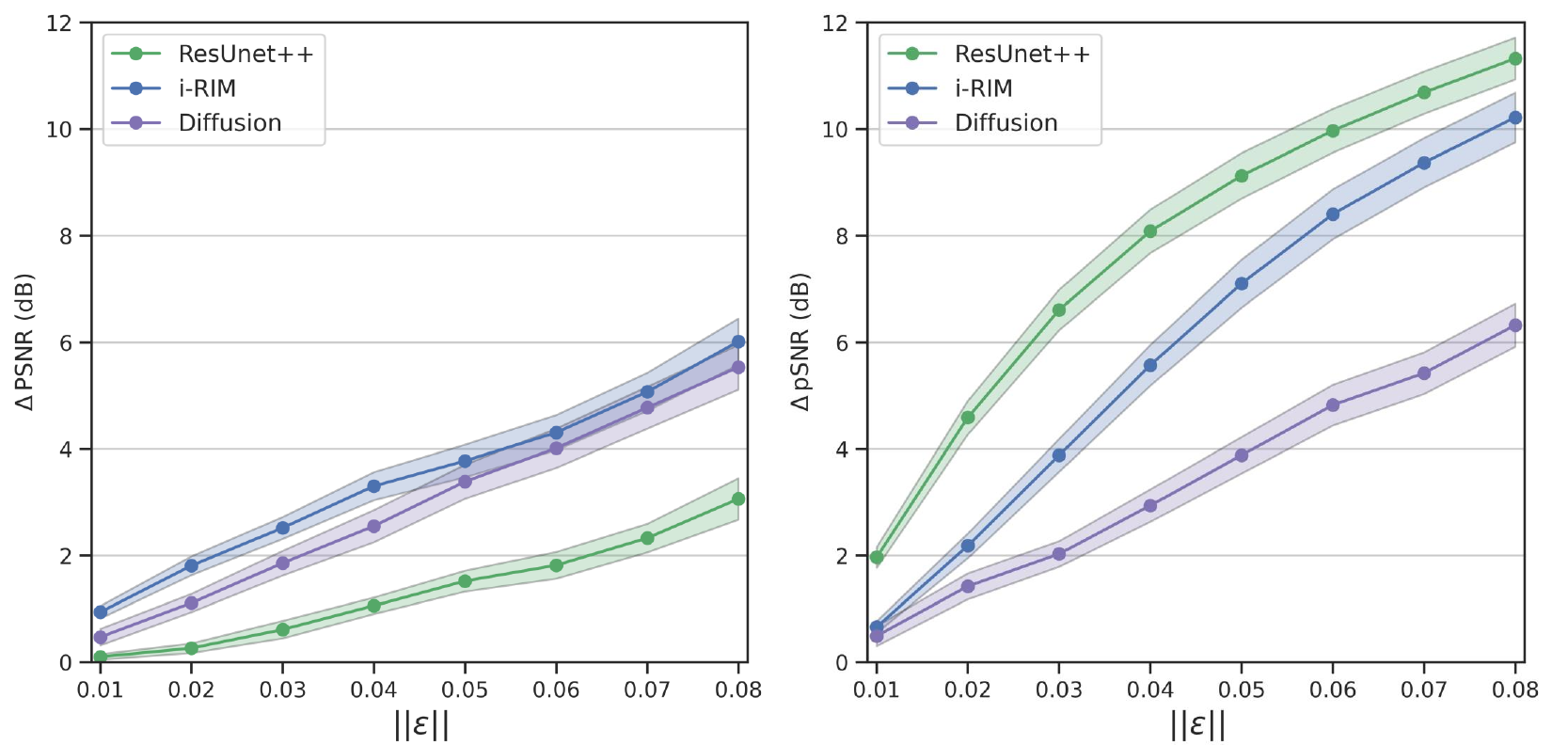}
  \caption{
    Comparisons of the pSNR change of all models against $l_2$-bounded adversaries with increasing perturbation amplitudes in the frequency space. 
    \textbf{left,} A worst-case perturbation was generated by using the trained i-RIM model.
    Comparisons in the left subplot correspond to our experiments outlined in Fig. \ref{fig:fig2}a.
    \textbf{Right,} A worst-case perturbation was generated by using the ResUnet++ model.
    Comparisons in the right subplot correspond to our experiments outlined in Fig. \ref{fig:fig2}b.
  }
  \label{fig:figs1}
\end{figure}

\begin{figure}[h!]
  \centering
  \includegraphics[trim=20 500 10 20, clip, width=\textwidth]{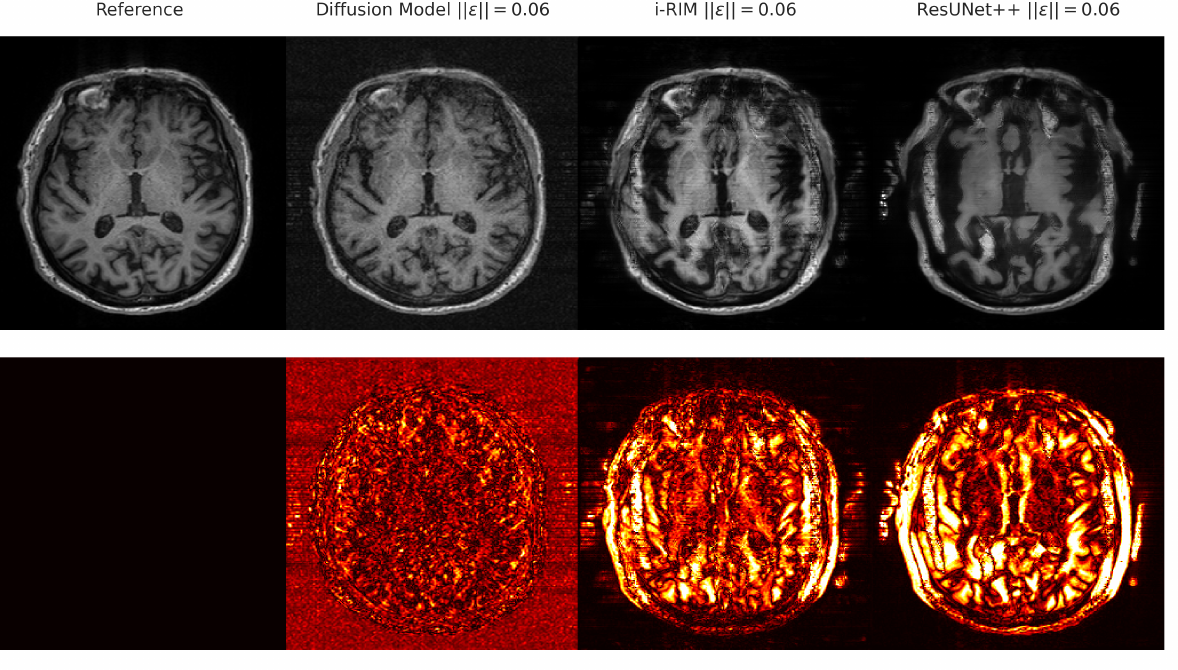}
  \caption{
    \textbf{Top,} Reconstructions of both supervised and unsupervised methods under the mid-level perturbation generated from ResUnet++. 
    \textbf{Bottom,} Mean absolute error map of all reconstructions. 
    The diffusion model, trained like a denoiser, exhibits an unstable reconstruction with a distinct noisy pattern that is difficult to distinguish from the clean reconstruction. 
  }
  \label{fig:figs2}
\end{figure}

\end{document}